\documentstyle[11pt,epsf]{article}
\newcommand{\be}{\begin{equation}}
\newcommand{\ee}{\end{equation}}
\newcommand{\bea}{\begin{eqnarray}}
\newcommand{\eea}{\end{eqnarray}}
\def\f{\frac}
\def\p{\partial}
\def\la{\lambda}
\def\om{\omega}
\def\ve{\varepsilon}

\def\le{\left}
\def\ri{\right}

\def\nn{\nonumber}

\def\ve{\varepsilon}
\def\t{\tau}
\def\s{\sigma}

\def\vp{\varphi}
\def\ka{\kappa}

\begin{document}

\begin{titlepage}

\vspace*{2cm}
\begin{center}
{\Large\bf Ground state energy of the modified 
Nambu--Goto string.} \\
\vskip 2cm
{\bf Leszek Hadasz}$  ^{ ^{\dag}}$

\vskip 0.5cm

Jagiellonian University, Institute of Physics \\
Reymonta 4, 30--059 Krak\'ow, Poland
\end{center}

\vskip 5mm

\begin{abstract}
We calculate, using zeta function regularization method,
semiclassical energy of the Nambu--Goto string 
supplemented with the boundary, Gauss--Bonnet term in the action
and discuss the tachionic ground state problem.

\end{abstract}

\vspace{\fill}

\noindent
\begin{tabular}{l}
TP-JU 12/97 \\
November 1997 \\
hep-th/9711101
\end{tabular}

\vskip 2cm
\noindent
\underline{\hspace*{10cm}}

\noindent
$ ^{\dag}$E--mail: hadasz@thp1.if.uj.edu.pl

\noindent
\end{titlepage}

It seems to exist a common belief that some string representation
of QCD would be very helpful in understanding  
such non--perturbative properties of quantum chromodynamics as 
the nature of the ground state or mechanism of confinement.
The conjecture of existence of such (at least approximate) description 
is supported by a number of facts \cite{intr6}, to mention only the nature
of the $1/N_c$ expansion \cite{intr1}, success of the dual models in
description of Regge phenomenology, area confinement law found
in the strong coupling lattice expansion \cite{intr3} or the existence
of flux--line solutions in confining gauge theories \cite{intr4,nkap3} and
the analytical results concerning two--dimensional QCD \cite{intr5}.

The  natural zeroth
order approximation in constructing the string description of QCD,
suggested by the approaches mentioned above, is
the Nambu--Goto string \cite{intr7}.
Unfortunately, when treated as a quantum system, it has many
well known drawbacks \cite{bnest,witt}, which include 
the non--physical dimension of the space--time
(D=26) or tachion and unwanted massless states in the
spectrum. 

In the free Nambu--Goto string model
the squared mass of the ground state is 
(up to some positive factor with the corresponding dimension
\cite{bnest})
just the Casimir energy
\cite{casen}. Therefore, revealing
the sign of this energy we can demonstrate the absence or persistence
of a tachion in the spectrum,
indicating in this way the
possibility of constructing a complete, consistent quantum theory
for the string model under consideration or invalidating 
such a construction.

A model which has been extensively investigated along this line
is the Nambu--Goto string with the point-like masses attached at the
ends \cite{nekla}. The results -- although not decisive for the
string propagating in the physical, four dimensional space-time --
are encouraging.

In this letter we want to consider another modification:
Nambu--Goto string with the boundary, Gauss-Bonnet
term added (for details see \cite{paw,lpaw}), defined by the action:
\be
\label{act}
S = \int\!\!d^2\xi\sqrt{-g}\left(-\gamma -\f{\alpha}{2}R\right),
\ee
where
\[
g_{ab} = \p_aX_\mu\p_bX^\mu,
\hspace{1cm}
a,b = \tau, \sigma,
\]
\[ g = \det(g_{ab}),
\]
$X_\mu$ gives immersion of the two--dimensional string
world--sheet parameterized by $(\tau,\sigma)$ into the four--dimensional
Minkowski spacetime, $\gamma$ and $\alpha$ are constants
and $R$ is the inner curvature of the
string world--sheet. 

The addition of the Gauss--Bonnet term is a
rather natural construction in the context of the effective QCD string.
The QCD string action should contain -- apart from the $X^\mu$ fields
-- also infinitely many fields describing for instance the transverse
shape of the chromoelectric flux joining the color sources. In constructing
the effective string action one integrates over such a fields and this
procedure inevitably leads to emergence of the intrinsic curvature term
in the action functional. Of course, it is then only the first one
out of the infinitely many terms with the growing number of derivatives.

In the four--dimensional space time (which we assume in this paper)
all the classical solutions of the model defined by the action functional
(\ref{act}) can be obtained (for details see \cite{bnest,paw}) by solving
the complex Liouville equation \cite{liouv}:
\be
\label{rl1}
\ddot{\Phi} - \Phi'' = 2q^2e^\Phi,
\ee
supplemented with the boundary conditions:
\be
\label{rl2}
e^\Phi = -\f1q\sqrt{\f{\alpha}{\gamma}}, \hspace{0.5cm}
\Im\;\Phi' = 0 \hspace{0.5cm} \mbox{for}
\hspace{0.5cm} \sigma = \pm\f{\pi}{2}.
\ee

The distinguished class of solutions of the Eqs. (\ref{rl1},\ref{rl2})
consists of static (i.e. $\t$--in\-de\-pen\-dent) Liouville fields. This solutions
are of the form:
\be
\label{ls0}
\Phi_0 = \log\le(\f{1}{q^2}\f{\la^2}{\cos^2\s}\ri) +
 i\pi,
\ee
where the parameters of the model $\alpha,\gamma$ and the 
parameters of the solution $\la,q$ are connected by the condition
\be
\label{lc}
\f{\cos^2\f{\pi\la}{2}}{\la^2} = \f{1}{q}\sqrt{\f{\alpha}{\gamma}},
\ee
following from Eq. (\ref{rl2}). The Liouville field (\ref{ls0}) corresponds
to the string which rotates rigidly in a plane:
\be
\label{rrr}
(X^\mu) = \f{q}{\la^2}\Big(\la\t,\;\cos\la\t\sin\la\s,
\;\sin\la\t\sin\la\s\;,0\Big).
\ee

In the paper \cite{lpaw} small oscillations around the solution (\ref{ls0})
have been investigated. If we write the full Liouville field $\Phi$ in
the form:
\be
\Phi = \Phi_0 + \Phi_1,
\ee
and assume that $\Phi_1$ is small, then we have:
\be
\label{lin}
\ddot{\Phi}_1 - \Phi_1'' = -V_\la(\sigma)\Phi_1,
\ee
together with the boundary conditions:
\be
\label{bc2}
\Phi_1 = 0; \hspace{5mm} \Im\;\Phi_1' = 0 \hspace{5mm}
\mbox{for} \hspace{5mm} \sigma = \pm\f{\pi}{2}.
\ee
Here we have defined:
\be
\label{potencjal}
V_\la(\sigma) = \f{2\la^2}{\cos^2\la\s}.
\ee

The general solution of the Eqs. (\ref{lin},\ref{bc2}) is of the form:
\be
\label{rozwstat}
\Phi_1(\t,\s)  =  
\sum_{n=1}^{\infty}D_n\cos\le(\om_n\t+\varphi_n\ri)
      \left\{\tan\la\sigma
      \cos\left(\om_n\sigma+\f{\pi n}{2}\right) 
      - \f{\om_n}{\la}\sin\left(\om_n\sigma+
          \f{\pi n}{2}\right)\right\},
\ee
where $D_n$ and $\varphi_n$ are arbitrary  constants and the
eigenfrequencies $\om_n$ are roots of the equation:
\be
\label{freq}
\om_n\tan\left[\f{\pi(\om_n+n)}{2}\right] =
 \la\tan\left(\f{\pi\la}{2}\right),
\ee
for $n\geq 1.$ The contribution to the string energy due to the
fluctuation $\Phi_1(\t,\s)$ is positive \cite{lpaw}. This shows
the classical stability of the rigidly rotating string against
small perturbations.

The constrains following from the reparametrization invariance of
the action functional (\ref{act}) reduce the number of the
functional d.o.f. of the $X^\mu$ field by two.
The boundary conditions, following from
the higher derivative terms in the action, impose additional restriction
which, in the case of considered small oscillations, (\ref{bc2}), forces 
the Liouville field (\ref{rozwstat}) to be real. 
This means that the 
string's functional d.o.f. are finally reduced to one. 
We can for instance describe the fluctuations of the string  
through the distance of the point with the
coordinate $\sigma$ form the center of rotation $\sigma = 0,$ 
evaluated at some fixed laboratory time $X^0,$
\bea
\label{length}
l(\tau,\sigma)  & = &  \f{q}{\la^2}\sin\la\s + \\
& + &    \f{2q}{\la^2}\sum_{n=1}^{\infty}
        D_n\le\{\f{\om_n^{+}}{\om_n^{-}}
       \cos\le(\om_n^{-}\s + \f{\pi n}{2}\ri) +
       \f{\om_n^{-}}{\om_n^{+}}
         \cos\le(\om_n^{+}\s+ \f{\pi n}{2}\ri) \ri\}
         \sin\le(\f{\la\om_n}{q}X^0 + \vp_n\ri) \nn
\eea
where $\om_n^\pm = \om_n\pm\la.$

Formula (\ref{length}) shows that oscillations of the weakly perturbed, 
rotating string can be expressed as a sum over non--interacting
modes with the frequencies
\be
\label{casfr}
\nu_n = \frac{\lambda\omega_n}{q}.
\ee

This suggest \cite{rajam} that in
the semiclassical (i.e. quadratic in the amplitudes of the oscillations)
approximation the quantum string Hamiltonian can be written as
\be
\label{strh}
H = E_0 + \f12\sum_{n=1}^{\infty}\nu_n
\left(b_n^{\dag}b_n + b_nb_n^{\dag}\right) = E_0 +
\sum_{n=1}^{\infty}\nu_nb_n^{\dag}b_n + \f12\sum_{n=1}^{\infty}\nu_n,
\ee
where
\be
\label{casen0}
E_0 = \f{\gamma q\pi}{2\la}\left(1+\f{\sin\pi\la}{\pi\la}\right)
\ee
is the classical string mass and the operator $b_n^{\dag}$ ($b_n$) creates 
(annihilates) the $n$--th mode of the
string oscillations.

The Hilbert space of the quantized string is now  constructed as a
Fock space generated by the creation operators $b_n^{\dag}$ in the
standard way \cite{rajam}. The only term that needs some care is the last
one in the formula (\ref{strh}).

To sum over mode frequencies which appear in the equation (\ref{strh}) 
is divergent and in order to give it a physical meaning we have to
adopt some regularization scheme. We choose to work with the $\zeta$
function regularization \cite{wiwib} and define
\be
\label{casen1}
E_{\rm C} \stackrel{\rm def}{=} \f12\lim_{s\to -1}\mu^{s+1}\zeta(s)
\ee
where, for $\Re\, s > 1,$
\be
\label{zetadef}
\zeta(s) = \sum_{n=1}^{\infty}\nu_n^{-s}
\ee
and the parameter $\mu$ with dimension of mass was introduced to
ensure that the r.h.s. of the expression (\ref{casen1}) has the
dimension of energy for arbitrary complex $s.$ The physically
interesting value $s=-1$ is obtained from (\ref{zetadef}) through
the analytic continuation.

In order to evaluate (\ref{casen1}) let us first 
rewrite the equation (\ref{freq}) for $\om_n$ 
in the equivalent form,
\be
D(\la,\om)  \stackrel{\rm def}{=}   \sin(\pi\om) \le[
\eta^2 - \f{1}{\om^2} - \f{2\eta}{\om}\cot(\pi\om)\ri] = 0,
\ee
where $\eta^{-1} = \la\tan\f{\pi\la}{2}.$ 
Using the standard methods of contour integration in the complex plane
we have
\be
\label{r6}
\sum_{n=1}^{N}\;\om_n^{-s} = 
\f{1}{2\pi i}\int_{{\cal C}_1}\!dz z^{-s} 
\f{d}{dz}\log D(\la,z),
\ee
where the integration contour ${\cal C}_1$ (Fig. 1) surrounds $N$ fist zeroes 
of the function $D.$ 

\centerline{\epsfbox{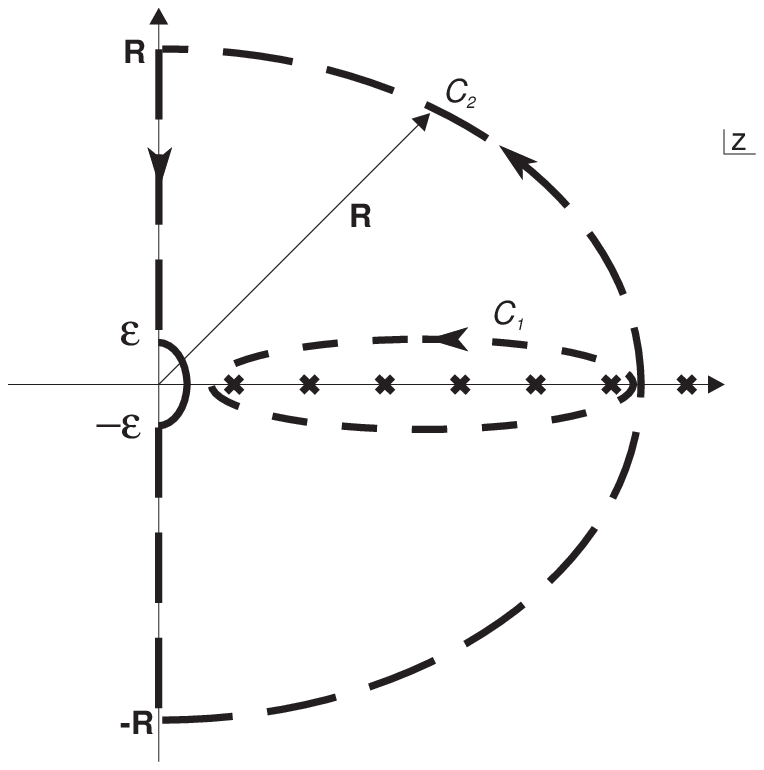}}

\centerline{Fig 1. The integration contours in the complex plane.}

Let us now  deform the integration contour ${\cal C}_1$ into 
${\cal C}_2$ which consists of a semi circle of radius $R,$ two line
intervals on imaginary axis $[i\ve, iR] \cup[-i\ve,-iR]$ and a semi circle
of radius $\ve.$ 
In the limit $R\to \infty$ the integral over 
the outer semi circle vanishes and we are left with the expression
\be
\label{casen2}
\sum_{n=1}^{\infty}\;\om_n^{-s} = 
\f{1}{\pi}\sin\le(\f{\pi s}{2}\ri)\le\{\f{\ve^{-s}}{s} + 
\int_{\ve}^{\infty}\!\!dx\;x^{-s}
\le[\pi\coth\pi x - \rho(\eta,x)\ri]\ri\},
\ee
where
$$
\rho(\eta,x) = 
\f{2}{x}\f{1+\eta x\coth(\pi x) + \eta\pi x^2{\rm cosech}^2(\pi x)}{
1 + 2\eta x\coth(\pi x) + \eta^2 x^2}.
$$

In order to be able to analytically continue r.h.s. of the equation
(\ref{casen2}) to the vicinity of $s=-1$ we subtract from the
integrand its (uniform in $\eta$) expansion
for $x \to \infty,$
\bea
\int_{\ve}^{\infty}\!\!dx\;x^{-s}
\le[\pi\coth\pi x - \rho(\eta,x)\ri]  &= &
\int_{\ve}^{\infty}\!\!dx\;x^{-s}
\le[\pi\coth\pi x - \rho(\eta,x)-\pi + \f{2}{x+\eta x^2}\ri]  +\nn
\\
&& \\
& + & \int_{\ve}^{\infty}\!\!dx\;x^{-s}
\le[\pi - \f{2}{x+\eta x^2}\ri]. \nn 
\eea
Defining
$$
F(\eta,s) = \int_{\ve}^{\infty}\!\!dx\; \f{x^{-s}}{x+\eta x^2}
$$
we have (for $\Re\;s > 1$)
$$
F(\eta,s) = \int_{\ve}^{\infty}\!\!dx\; x^{-(s+1)}\le[1- 
\eta\f{x}{1+\eta x}\ri] = \f{\ve^{-s}}{s} - \eta F(\eta,s-1),
$$
or equivalently
$$
F(\eta,s) = \f{1}{\eta}\le[\f{\ve^{-(s+1)}}{s+1} - F(\eta,s+1)\ri].
$$
This relations allows us to analytically continue the function
$F(\eta,s)$ to the vicinity of $s = -1$ with the result
$$
F(\eta,s) \stackrel{s\to -1}{=} \f{1}{\eta}\f{1}{s+1}  + {\cal O}(s+1).
$$

Analytic continuation of the integral
$$
\int_{\ve}^{\infty}\!\!dx\;\pi x^{-s}
$$
gives zero and we finally arrive at the expression
\be
\label{regsum}
\sum_{n=1}^{\infty}\;\om_n^{-s} \stackrel{s\to -1}{=}
\f{2}{\pi\eta}\f{1}{s+1} - \f{1}{12} 
+ \f{1}{\pi}\int_{0}^{\infty}\!\!dx\;\le[x\rho(\eta,x)- \f{2}{1+\eta x}\ri] +
{\cal O}(s+1).
\ee

Inserting (\ref{regsum}) into the definition (\ref{casen1}) we get
\be
\label{casen3}
E_{\rm C} {=} \lim_{s\to -1}\le[
  \f{\la}{q}\le\{\f{1}{\pi\eta}\f{1}{s+1} + 
  \f{1}{\pi\eta}\log\tilde\mu + 
  \f{1}{2\pi}\int_{0}^{\infty}\!\!dx\;\le[x\rho(\eta,x)- \f{2}{1+\eta x}\ri] -
  \f{1}{24}\ri\} + {\cal O}(s+1)\ri],
\ee
where $\tilde\mu = \mu q/\la$ is also an arbitrary (due to arbitrariness
of $\mu$), but now dimensionless constant.

Because of the pole in the first term the Casimir energy (\ref{casen3}) is
a divergent quantity and we
need to apply a renormalization procedure to remove this divergence.

Let us first notice that the effective action of a single component, 
free field, which describe the only functional
d.o.f. of the string oscillations left after the gauge fixing and 
imposition of the boundary conditions, has to be of the form:
\be
\label{efact}
S_{\rm eff} = -\int_{\t_1}^{\t_2}\!\!d\t
               \int_{-\f{\pi}{2}}^{\f{\pi}{2}}\!\!d\s
	\sqrt{-g^{(0)}}\left\{\f12g^{(0)}{}^{ab}\p_a\psi\p_b\psi -
	\f12\le(m^2 + \xi R^{(0)}\ri)\psi^2\ri\},
\ee
where $a,b=\t,\s,$ 
$$
g^{(0)}_{ab} = \f{q^2}{\la^2}\cos^2(\la\s)
$$
is the induced metric tensor on the rotating string worldsheet, 
$R^{(0)}$ is the 
induced curvature scalar and $\xi,m^2$ are constants. Choosing 
$m=0, \xi = 1,$ we get from (\ref{efact}) the equation of motion 
which coincides with the equation (\ref{lin}).
Moreover, the quantum Hamiltonian which follows from the action
functional (\ref{efact}),
\be
\label{efham}
\hat H_{\rm eff} = \f12 \int_{-\f{\pi}{2}}^{\f{\pi}{2}}\!\!d\s
      \; \le[\hat\pi^2(\t,\s) + \le(\p_\s\hat\psi(\t,\s)\ri)^2 + 
	V(\s)\hat\psi^2(\t,\s)\ri],
\ee
 differs from the Hamiltonian
(\ref{strh}) only by a factor $\f{q}{\la}$ following from the fact
that (\ref{efham}) generates translation in the worldsheet time
$\tau$ and (\ref{strh}) in the laboratory time $X^0 = \f{q}{\la}\t,$
and by a constant term given by the background field energy
$E_0$ (\ref{casen0}).

Because the Hamiltonian (\ref{efham}) contains products of quantum fields
in the same space--time point, the renormalization
procedure has to be applied to it to render its eigenvalues finite.
This can be  accomplished, \cite{rajam},
 by replacing the Hamiltonian $\hat H_{\rm eff}$ 
by its normal-ordered form $:\hat H_{\rm eff}:.$ 

The normal ordered field product can be written in the form:
$$
:\hat\psi^2: = \hat\psi^2 + C
$$
where $C$ is a counter term. Now, the factor in front of the pole term
in (\ref{casen3}) is of the form
$$
\f{1}{\pi\eta} = \f{1}{4\pi}\int_{-\f{\pi}{2}}^{\f{\pi}{2}}\!\!d\s\;V(\s).
$$
This shows that the replacement
\be
\label{repl}
\hat\psi^2 \to :\hat\psi^2: = \hat\psi^2 - \f{1}{2\pi}\f{1}{s+1}
\ee
in the Hamiltonian (\ref{efham}) renders its eigenvalues finite. Let
us also note that in the limit of flat background
the applied procedure is the usual subtraction of the 
(infinite) constant from the quantum Hamiltonian, which is equivalent
to shifting the origin of the energy scale.

After eliminating the divergent term the Casimir energy
(\ref{casen3}) is still not uniquely determined
due to the dependence on the
arbitrary parameter $\tilde\mu.$ This can be of course traced back
to the renormalization prescription (\ref{repl}): without
spoiling the condition of finiteness of the Hamiltonian eigenvalues we 
could add to the r.h.s. of (\ref{repl}) arbitrary
constant $\log\tilde\mu.$ 

The actual value of $\tilde\mu$ cannot be
calculated in the presented framework --- it is an additional
parameter that appears in the model at the quantum level and
in order to determine it we need some experimental data.
Le us note however that the dependence of the Casimir energy 
on the quantum scale $\tilde\mu$ is, in fact, not surprising.
The Casimir energy is intimately related to the one--loop physics
\cite{wiwib}, and the occurrence of anomalous scale dependence 
in one--loop field theory is by now a well--understood phenomenon
\cite{ra,itz}

In order to be able to see the consequences of the formula (\ref{casen3})
with divergent term being subtracted let us note that it can be
rewritten as a function of the unperturbed string length,
$$
L = \f{2q}{\la^2}\sin\le(\f{\pi\la}{2}\ri),
$$
and a single dimensionless parameter $\ka = \sqrt{\f{\alpha}{\gamma L^2}}:$
\be
\label{casen4}
E_{\rm C}(\ka) =
  \f{2}{\pi L}\;\le\{
  \f{\sin^2\le(\f{\pi\la(\ka)}{2}\ri)}
  {\cos\le(\f{\pi\la(\ka)}{2}\ri)}\log\tilde\mu +
  \f{\sin\le(\f{\pi\la(\ka)}{2}\ri)}{2\la(\ka)}
  \int_{0}^{\infty}\!\!dx\;\le[x\rho(\eta(\ka),x)- 
  \f{2}{1+\eta(\ka) x}\ri] -
  \f{1}{24}\ri\}
\ee
where
$$
\sin\le(\f{\pi\la(\ka)}{2}\ri)= \sqrt{1+\ka^2}-\ka,
\hskip 1cm
\eta(\ka) = \le[\la(\ka)\tan\le(\f{\pi\la(\ka)}{2}\ri)\ri]^{-1},
$$
and we have used the relation (\ref{lc}).

The factor in front of the undetermined
term $\log\tilde\mu$ diverges for $\alpha\to 0$ 
(although rather slowly, as $\alpha^{-\f14}$).
Consequently, if we demand that our model should give finite
value of the Casimir energy in the Nambu--Goto limit 
we have the unique choice $\log\tilde\mu = 0.$ The values of the
Casimir energy computed in this way,
plotted on the Fig.2 below, are negative for all values of the 
parameters $\alpha, \gamma$
and for any length $L$ of the unperturbed, rotating
string. Consequently, our model does not possess the ground state --
we can always lower the energy by taking strings with smaller
length $L.$ This result is true for all $\tilde\mu < 1$ where the
additional term $\sim\log\tilde\mu$ lowers the energy further.

\centerline{\epsfbox{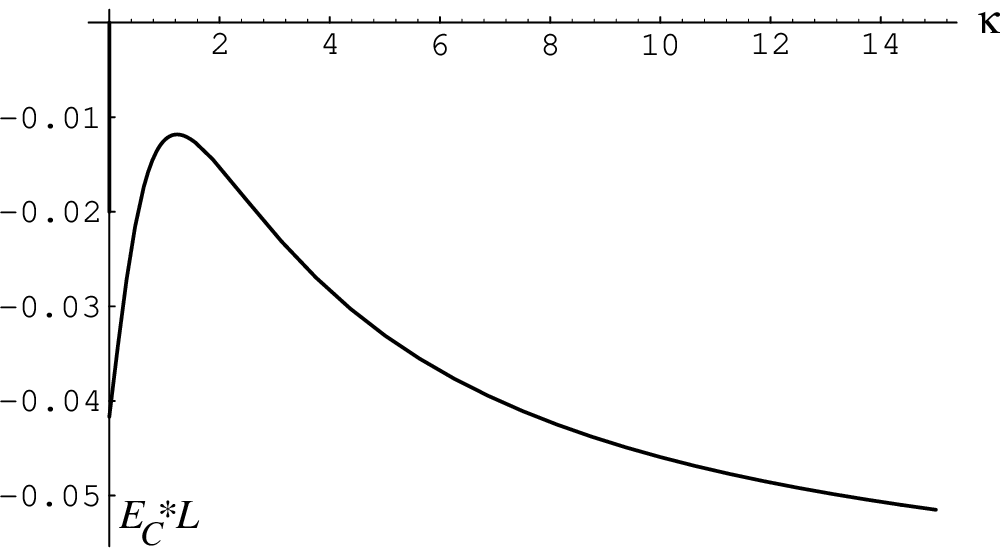}}

\vskip -1cm
\begin{center}
Fig 2. The Casimir energy of the NGGB model as a function
of the dimensionless parameter
$\ka = \sqrt{\f{\alpha}{\gamma L^2}}$ for $\tilde\mu = 1.$
\end{center}

\newpage

In the Nambu--Goto ($\kappa = 0$)  limit formula (\ref{casen4})
gives
$$
E_{\rm C}(0) = -\f{1}{12}\f{1}{L}.
$$
This is different from the celebrated L\"usher term \cite{lusher},
$$
E_{C}^{\rm L} = -\f{\pi}{12}\f{1}{L},
$$
but the reasons are obvious. First, L\"usher term is derived for the
string with fixed ends and the oscillation frequencies equal
$$
\nu_{n}^{\rm L} = \f{\pi n}{L},
$$
while in our, rotating string case we have (\ref{casfr})
$$
\nu_n(\kappa = 0) = \f{2n}{L}.
$$
Second, in considered model we have only planar oscillations
and this gives additional factor 1/2. This agreement can be viewed
as a consistency check for the used regularization method. 

If the quantum mechanically generated scale $\tilde\mu > 1,$
the conclusions concerning the existence of the ground state 
are different. The first, positive term in the
Eq. (\ref{casen4}) grows indefinitely for $\alpha\to 0$ and it is
always possible to choose the parameters of the model in such
a way that the Casimir energy is positive. Consequently, the semiclassical
(i.e. classical + Casimir) string energy,
$$
E_{\rm sc} =   c_2 L + \f{c_2}{L},
$$
where $L$ is the string length and the positive constants 
$c_1,$ $c_2$ can be read off from Eqs. (\ref{casen0}) and (\ref{casen4})
respectively, possesses positive minimum for some non-zero
value of the rotating string length. This configuration can 
be taken as a correct starting point for the semiclassical quantization
procedure.

\vskip 3mm

\noindent
{\Large\bf Acknowledgments.}

I would like to thank Prof. Henryk
Arod\'z, Dr Pawe\l{} W\c{e}grzyn and Dr Jacek Dziarmaga
for numerous helpful discussions.
This work is partially supported by the KBN grant 2 P03B 095 13.

\end{document}